\documentclass[11pt]{cernrep}
\usepackage{graphicx}
\usepackage{here}

\newcommand{\beq}{\begin{equation}}
\newcommand{\eeq}{\end{equation}}
\newcommand{\beqa}{\begin{eqnarray}}
\newcommand{\eeqa}{\end{eqnarray}}
\newcommand{\gsim}{\buildrel > \over {_\sim}}
\newcommand{\lsim}{\buildrel < \over {_\sim}}

\newcommand{\gev}{{\rm ~GeV}}
\newcommand{\ie}{{\it ie.}}

\newcommand{\as}{\alpha_{s}}
\newcommand{\lqcd}{\Lambda_{QCD}}
\newcommand{\qpair}{Q\bar Q}

\newcommand{\jpsi}{J/\psi}
\newcommand{\order}[1]{${\cal O}(#1)$}
\newcommand{\morder}[1]{{\cal O}(#1)}

\begin{document}

\title{
\hbox to\hsize{\normalsize\hfil\rm HIP-2002-38/TH}
\hbox to\hsize{\normalsize\hfil\rm LAPTH-930/02}
\hbox to\hsize{\normalsize\hfil hep-ph/0209365}
\hbox to\hsize{\normalsize\hfil September 30, 2002}%\protect\today} 
\vskip 40pt
%\title{
\centerline{COMOVER ENHANCEMENT OF QUARKONIUM PRODUCTION\footnote{To
    appear in the proceedings of the CERN 2001-2002  
workshop `Hard Probes in Heavy Ion Collisions at the LHC'.}}}
\author{P. Hoyer$^{1}$,
N. Marchal$^{2}$,
S. Peign\'e$^{2}$}
\institute{$^{1}$Department of Physical Sciences and Helsinki 
Institute of Physics\\
POB 64, FIN-00014 University of Helsinki, Finland\\
$^{2}$LAPTH, Chemin de Bellevue B.P.110, 
F-74941 Annecy-le-Vieux Cedex, France}
\maketitle
\begin{abstract} 
Quarkonium data
suggest an {\em enhancement} of the hadroproduction
rate from interactions of the heavy quark pair with a comoving color
field generated in the hard $gg \to \qpair$ subprocess. We review the
motivations and principal consequences of this comover enhancement
scenario (CES). 
\end{abstract}

\section{THE QUARKONIUM THERMOMETER}

The production of heavy $\qpair$ quarkonia has the potential of
offering valuable insights into QCD dynamics, complementary to those
given by open heavy flavor production. In both cases, the creation of
the heavy quark pair requires an initial parton collision of hardness
\order{m_Q}. Most of the time the heavy quarks hadronize
independently of each other and are incorporated into separate
hadrons. The QCD factorization theorem exploits the conservation of
probability in the hadronization process to express the total heavy
quark production cross section in terms of target and projectile
parton distributions and a perturbative subprocess cross section such
as $\sigma(gg\to \qpair)$.

The quarkonium cross section is a small fraction of the open flavor
one and is thus not constrained by the standard QCD factorization
theorems \cite{Collins:gx}. 
Nevertheless, it is plausible that the initial creation of
the heavy quarks is governed by the usual parton distributions and
hard subprocess cross sections, with the invariant mass of the
$\qpair$ pair constrained to be close to threshold. Before the
quarkonium emerges in the final state there can, however, be further
interactions which, due to the relatively low binding energy, can
either ``make or break'' the bound state. Quarkonium studies can thus
give new information about the environment of hard production, from
the creation of the heavy quark pair until its ``freeze-out''. The
quantum numbers of the quarkonium state furthermore impose
restrictions on its interactions. Thus states with negative charge
conjugation ($\jpsi$, $\psi'$) or total spin $J=1$ ($\chi_{c1}$)
require the $\qpair$ pair to interact at least once after its
creation via $gg\to \qpair$.

Despite an impressive amount of data on the production of several
quarkonium states with a variety of beams and targets we still have a
poor understanding of the underlying QCD dynamics. Thus quarkonia
cannot yet live up to their potential as `thermometers' of $A+B$
collisions, where $A,B = \gamma^{(*)}$, hadron or nucleus. Rather, it
appears that we need simultaneous studies and comparisons of several
processes to gain insight into the production dynamics.

We will now summarize the successes and failures of the
Color Singlet Model \cite{Berger:1980ni,Baier:1981uk},
which we consider as a guideline for understanding the
nature of the quarkonium production dynamics.

\section{SUCCESSES AND FAILURES OF THE COLOR SINGLET MODEL} \label{csmsect}

In the Color Singlet Model (CSM), the $\qpair$ pair is directly prepared
with the proper quantum numbers in the initial hard subprocess and
further interactions are assumed to be absent. The quarkonium production
amplitude is then given by the overlap of the non-relativistic wave
function with that of the $\qpair$ pair.

This model at NLO correctly predicts the
normalization and momentum dependence of the $\jpsi$ photoproduction
rate \cite{Kramer:2001hh,Adloff:2002ex}. While the absolute
normalization of the CSM prediction is uncertain by a factor of $2 -
3$ there appears to be no need for any additional production
mechanism, for longitudinal momentum fractions $0.3 \leq x_F \leq 0.9$
and $1 \leq p_\perp^2 \leq 60\gev^2$. The comparison with
leptoproduction data \cite{Adloff:2002ey} is less conclusive since
only LO CSM calculations exist.

The CSM underestimates the directly produced $\jpsi$ and
$\psi'$ hadroproduction rates by more than an order of magnitude.
This is true both at low $p_{\perp} \lsim m_c$ (fixed target)
\cite{Kaplan:1996tb} and at high
$p_{\perp} \gg m_c$ (collider)  \cite{Kramer:2001hh} data. Similar
discrepancies for the $\Upsilon$
states~\cite{Abe:1995an,Alexopoulos:1995dt,Affolder:1999wm}
indicate that the anomalous
enhancement does not decrease quickly with increasing quark mass.

The
inelastic cross section ratio $\sigma(\psi')/\sigma_{\rm dir}(\jpsi)$ is
similar in photoproduction \cite{Bertolin:1999yj} and hadroproduction
\cite{Alexopoulos:1997yd,Lourenco:1996wn} and consistent with the value
$\simeq 0.24$ expected in the CSM~\cite{Vanttinen:1994sd}.
The ratio does not depend on $x_F$
in the projectile fragmentation region and is independent of the
nuclear target size in $hA$ collisions. The CSM thus underestimates
the $\jpsi$ and $\psi'$
hadroproduction cross sections, as well as that of the
$\chi_{c1}$ \cite{Vanttinen:1994sd},
by similar large factors. The quantum numbers of these charmonium
states require final-state gluon emission in the CSM, $gg \to \jpsi g$.
This emission is not required for the $\chi_{c2}$ where the 
CSM cross section $\sigma(gg \to \chi_{c2})$ is only a factor $\sim 2$
below the hadroproduction data \cite{Vanttinen:1994sd}.

In the CSM, $\chi_c$ photoproduction is suppressed by a power of $\alpha_s$
compared to the $J/\psi$ and $\psi'$ production rates.
One indeed observes a smaller value
of the $\sigma (\chi_{c2})/\sigma ( J/ \psi)$ ratio in
photoproduction~\cite{Roudeau:1988wb} than in
hadroproduction~\cite{Antoniazzi:1992iv,Antoniazzi:1993yf}.

\section{THE COMOVER ENHANCEMENT SCENARIO (CES)}

The analysis of agreements and discrepancies between the CSM and quarkonium
data led to the comover enhancement scenario of quarkonium
production~\cite{Hoyer:1998ha,Hoyer:1998dr}.
Hadroproduced $\qpair$ pairs are created within a comoving color field
and form $\jpsi$, $\psi'$ and $\chi_{c1}$ through gluon {\em
absorption} rather than emission, enhancing the cross
section relative to the CSM since the pair gains rather than loses
energy and momentum. The $\chi_{c2}$ cross section is not as strongly 
influenced since no gluon needs to be absorbed or emitted. Most importantly,
such a mechanism
is consistent with the success of the CSM in photoproduction
since no color fields are expected in the photon fragmentation region,
$x_F \gsim 0.3$.

\begin{figure}[thb]
\begin{center}
\includegraphics[width=6cm]{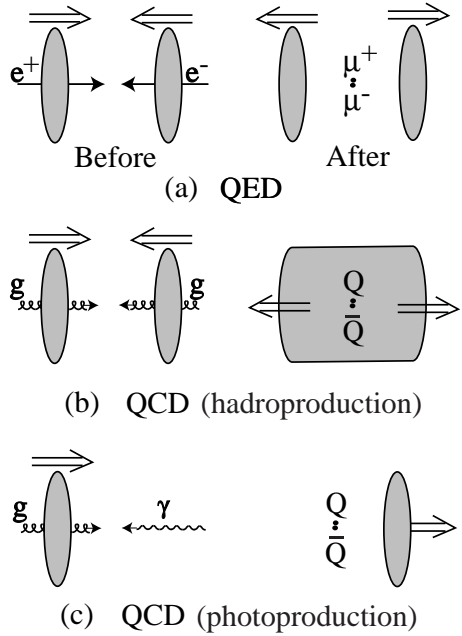}
\caption{Schematic scenarios of gauge field interactions are compared:
(a) $e^+ e^- \rightarrow \mu^+ \mu^-$ in QED; (b) hadroproduction of a 
$Q \overline Q$ pair, {\it e.g.}\ $gg \rightarrow Q \overline Q$; and (c)
photoproduction of a $Q \overline Q$ pair, $\gamma g \rightarrow Q \overline
Q$.  The creation of a comoving color field is specific to hadroproduction,
(b).}
\label{comofield}
\end{center}
\end{figure}

The effect of a comoving color field is illustrated in Fig. \ref{comofield}. 
Light charged particles carry
gauge fields which are radiated in high energy annihilations
into a heavy particle pair. In $e^+e^- \to \mu^+\mu^-$ annihilations, the
photon fields pass through each other and materialize as forward
bremsstrahlung, Fig. \ref{comofield}(a). 
In $gg \to \qpair$, on the other hand, the
self-interaction of the color field can also result in the creation of a
gluon field at intermediate rapidities, Fig. \ref{comofield}(b). Hadroproduced
$\qpair$ pairs thus find themselves surrounded by a color field. We
postulate that interactions between the $\qpair$ pair and this
comoving field are important in quarkonium hadroproduction. In direct
photoproduction, the incoming photon does not carry any color field
and the $\qpair$ pair is left in a
field-free environment after the collision, Fig. \ref{comofield}(c). 
The proposed rescattering thus does not affect photoproduction.

The importance of rescattering effects in hadroproduction as compared
to photoproduction is also suggested by data on open charm
production. Hadroproduced $D \bar D$ pairs are almost uncorrelated in
azimuthal angle \cite{Aitala:1998kh}, at odds with
standard QCD descriptions. Photoproduced pairs on the other hand,
emerge nearly back-to-back \cite{Frabetti:1996vi}, following the
charm quarks of the underlying $\gamma g\to c\bar c$ process.

Since the hardness of the gluons radiated in the
creation process increases with quark mass, the rescattering effect
persists for bottomonium. Due to the short timescale of the radiation
the heavy quark pair remains in a compact configuration during
rescattering and overlaps with the quarkonium wave function at the
origin. The successful CSM result for
$\sigma(\psi')/\sigma_{\rm dir}(\jpsi)$~\cite{Vanttinen:1994sd} 
is thus preserved.

The $\qpair$ pair may also interact with the more distant projectile
spectators after it has expanded and formed quarkonium. Such
spectator interactions are more frequent for nuclear projectiles and
can cause the breakup (absorption) of the bound state. This
conventional mechanism of quarkonium suppression in nuclei is thus
fully compatible with, but distinct from, 
interactions with the comoving color field.

We have investigated the consequences of the CES using pQCD to
calculate the interaction between the $\qpair$ and the comoving
field. While we find consistency with data, quantitative predictions
depend on the structure of the comoving field. Hence tests of the CES 
must rely on its generic features which we discuss below.

\section{GENERIC FEATURES}

The CES distinguishes three proper timescales in quarkonium production:
\begin{itemize}
\item $\tau_Q \sim 1/m_Q$, the $\qpair$ pair production time;
\item $\tau_{AP}$,
the DGLAP scale over which the comoving field is created and
interacts with the $\qpair$ pair;
\item $\tau_\Lambda \sim 1/\lqcd$, while rescattering with
comoving spectators may occur.
\end{itemize}
In the following we will consider quarkonium production
at $p_{\perp} \lsim m_Q$.
In quarkonium production at $p_\perp \gg m_Q$, a large $p_\perp$
parton is first created on a timescale $1/p_\perp$, typically through
$gg \to gg^*$. The virtual gluon then fragments, $g^* \to \qpair$, in
proper time $\tau_Q$.
Thus high $p_\perp$ quarkonium production is also based on the CES 
\cite{Peigne:2000wd}.

\subsection{Timescale $\tau_Q \sim 1/m_Q$: creation of the $\qpair$ pair}

The $\qpair$ pair is created in a standard parton subprocess,
typically $gg \to \qpair$, at a time scale $\tau_Q \sim 1/m_Q$.
This first stage is common to other
theoretical approaches such as the Color Evaporation Model 
(CEM)~\cite{Fritzsch:1977ay,Halzen:1977rs,Gluck:1977zm} and non-relativistic
QCD~\cite{Bodwin:1994jh,Beneke:1996tk}, here referred to as
the Color Octet Model (COM). The 
momentum distribution of the
$\qpair$ is determined by the product of projectile $(A)$ and target
$(B)$ parton distributions, such as
$g_A(x_1) g_B(x_2)$ where $A,B$
may be a lepton, photon, hadron or nucleus. This production process
is consistent with the quarkonium data.

According to pQCD the $\qpair$ is dominantly produced in a color
octet, $S=L=0$ configuration, close to threshold. Such a state can
obtain the quarkonium quantum numbers through a further interaction which
flips a heavy quark spin and turns the pair into a color singlet. The
amplitude for processes of this type are suppressed by the factor
$k/m_Q$ where $k$ is the momentum scale of the interaction. The
various theoretical approaches differ in the scale assumed for $k$.
\begin{description}
\item{CSM:} Here $k=\morder{m_Q}$. Thus $\jpsi$
production proceeds via the emission of a hard gluon in the primary
process, $gg \to \qpair+g$. The $\chi_{c2}$ is produced
without gluon emission, 
$gg \to \chi_{c2}$, through a subdominant $S=L=1$ color singlet
production amplitude.
\item{COM:} The $\qpair$ quantum numbers are
changed via gluon emission at the bound state momentum scale
$k=\morder{\as m_Q}$. This corresponds to an expansion in powers of
the bound state velocity $v=k/m_Q$, introducing nonperturbative
matrix elements that are fit to data.
\item{CEM:} Here $k=\morder{\lqcd}$. Soft
interactions are postulated to change the $\qpair$ quantum numbers
with probabilities that are specific for each quarkonium state but
independent of kinematics, projectile and target.
\item{CES:} The quantum numbers of the $\qpair$ are changed
in perturbative interactions with a comoving field at scale
$k=\morder{1/\tau_{AP}}$, as described below.
\end{description}

\subsection{Timescale $\tau_{AP}$: interactions with the comoving field}

The scale $\tau_{AP}$ refers to the time in which collinear
bremsstrahlung, the source of QCD scaling violations, is emitted in
the heavy quark creation process \cite{Hoyer:1998ha}.
Thus $1/\tau_{AP}$ characterizes
the effective hardness of logarithmic integrals of the type
$\int_{\mu_F}^{m_Q} dk/k$ where $\mu_F \ll m_Q$ is the factorization
scale. We stress that $1/\tau_{AP}$ is an intermediate but still
perturbative scale, $\tau_Q \ll \tau_{AP} \ll \tau_{\Lambda}$, which
grows with $m_Q$.

The fact that the $\qpair$ pair acquires the quarkonium quantum
numbers over the per\-tur\-ba\-ti\-ve time\-scale $\tau_{AP}$ is a feature
of the CES and distinguishes it from other approaches. At this
time, the pair is still compact and couples to quarkonia
via the bound state wavefunction at the origin
or its derivative(s). Thus no
new parameters are introduced in this transition. However, the
interactions of the $\qpair$ pair depend on the properties
of the comoving color field such as the intensity and polarization.
Quantitative predictions in the CES are only possible
when the dependence on the comoving field is weak.

Ratios of radially excited quarkonia, such as
$\sigma(\psi')/\sigma_{\rm dir}(\jpsi)$, are insensitive to the comoving
field and are thus expected to be process-independent when
absorption on spectators at later times can be ignored, see below.
The fact that this ratio is observed to be roughly universal
\cite{Alexopoulos:1997yd,Lourenco:1996wn} is one of the
main motivations for the CES. Even the measured variations of the
ratio in different reactions agree with expectations, see 
Ref.~\cite{Hoyer:fq} for a discussion of its
systematics in elastic and inelastic photoproduction, leptoproduction
and hadroproduction at low and high $p_\perp$.

The ratio $\sigma(\chi_{c1})/\sigma(\chi_{c2})$ is
measured to be $0.6 \pm 0.3$ in pion-induced~\cite{Koreshev:1996wd}
and $0.31 \pm 0.14$ in proton-induced~\cite{Alexopoulos:1999wp}
reactions. The CSM underestimates
this ratio~\cite{Vanttinen:1994sd} as well
as that of~$ \sigma ( J/ \psi )/\sigma (\chi_2 )$~\cite{Vanttinen:1994sd}.
The rescattering contribution
increases $\sigma ( J/\psi )$ and $\sigma (\chi_1 )$, enhancing
the above ratios. 

\subsection{Nuclear target dependence}

The quarkonium cross section can be influenced by rescattering effects
in both the target and projectile fragmentation regions. For
definiteness, we assume the charmonium is produced in the
projectile fragmentation region, $x_F > 0$.

The nuclear target dependence is usually parametrized as $\sigma(hA \to
\jpsi+X) \propto A^\alpha$. It turns out that  $\alpha < 1$ and
obeys Feynman scaling, \ie, $\alpha$ decreases with~$x_F$ rather
than with the momentum fraction~$x_2$ of the target
parton~\cite{Hoyer:1990us,Leitch:1999ea}. The comparison with lepton pair
production in the Drell-Yan process shows that the $\jpsi$ nuclear suppression
cannot be attributed to shadowing of parton distributions in the
nucleus~\cite{Lourenco:1996wn}. The $A$-dependence is thus difficult 
to explain in the CSM, COM and
CEM approaches.

In the Feynman scaling regime, we may assume that the $\qpair$ pair
energy is high enough to remain compact while traversing the target.
The relative transverse momentum of the $Q$ and $\bar Q$ could
increase as a result of rescattering in the target, thus suppressing
the binding probability. However, this
explanation is unlikely in view of the absence of nuclear
suppression in photoproduction~\cite{Amaudruz:1991sr}.

In the CES, the nuclear target suppression is ascribable
to absorption of the comoving color field in the target nucleus. This
field is emitted by a projectile parton with transverse size
$\tau_{AP}$, larger than the size, $\sim 1/m_Q$, of the
$\qpair$ pair. Due to Lorentz time dilation, the field is emitted long
before reaching the target and reinteracts with the $\qpair$ long after 
passing the
target. Absorption of the comoving field in the target implies
suppression of $\jpsi$ production in the CES. At high energies,
we have $x_1 \simeq x_F$, which explains the Feynman scaling of this
effect.  Moreover, as $x_F$ increases,
less energy is available to be radiated to the gluon field which
therefore becomes softer, further increasing its absorption in the
target and thus the nuclear suppression.

This explanation is consistent with the fact that the nuclear target
suppression of the $\jpsi$ and the $\psi'$ is found to be the same
for $x_F \gsim 0.2$~\cite{Leitch:1999ea}. It also predicts that the 
suppression will be
similar for $\chi_{c1}$ production. On the other hand, the nuclear
target suppression should be {\em reduced} for $\chi_{c2}$ since a
substantial fraction is directly produced 
without a comoving field. A measurement of
$\sigma(hA \to \chi_{c}+X)$ in the projectile fragmentation region
would thus constitute an important test of the CES.

In a Glauber picture of the nuclear suppression, a relatively large value
for the absorption cross section is required, 
$\sigma_{\rm abs} \sim 5 \ {\rm mb}$.
We interpret this value as the joint cross section of the $\qpair$ pair
and the comoving field, thus of order $\tau_{AP}^2 \gg 1/m_Q^2$.
Since $1/\tau_{AP}$ scales with $m_Q$,
we expect less nuclear absorbtion for the
$\Upsilon$ states than for the $J/ \psi$, as observed
experimentally~\cite{Alde:1991sw,Leitch:di}.

\subsection{Timescale $\tau_\Lambda \sim 1/\lqcd$: interactions with
comoving spectators}

By the time, $\tau_\Lambda \sim 1$ fm, that the $\qpair$ pair encounters
comoving projectile spectators, the pair has already expanded and is
distributed according to the quarkonium wave function. The spectator
rescattering effects are thus independent of the quarkonium formation
process. Larger and more loosely bound charmonia are
more easily broken up by secondary scattering. Hence the $\chi_c$
and $\psi'$ cross sections should be depleted compared to that of the
$\jpsi$. Likewise bottomonium is generally less affected by spectator
interactions than charmonium.

Spectator interactions at large enough $x_F$
are likely to be unimportant for hadron
projectiles judging from the approximate universality of the
$\sigma(\psi')/\sigma_{\rm dir}(\jpsi)$ ratio in photo- and
hadroproduction. The lower ratio seen in nucleus-nucleus
collisions~\cite{Abreu:1999nn},
on the other hand, is most naturally explained by absorption on
spectators. It would be important to confirm this by also measuring the
ratio in $Ah$ scattering.

\section{SUMMARY}

Data on quarkonium production have proved challenging for QCD models.
The richness of the observed phenomena indicates that quarkonium cross
sections are indeed sensitive to the environment of the hard QCD
scattering. We may, however, only be able to decipher its message
through systematic experimental and theoretical studies of several
species of quarkonia produced with a variety of beams and targets in
a range of kinematic conditions.

It is thus useful to begin by asking fairly general questions
about the production mechanism before fitting detailed models to
data. This is the spirit of the comover enhancement scenario
that we have discussed here, motivated by apparent regularities of
the data and general theoretical constraints.

\vskip1cm
\noindent
\section*{ACKNOWLEDGEMENTS} 
S.~P. would like to thank Cristina Volpe for very stimulating
discussions. We also thank Ramona Vogt for her careful reading of our 
manuscript.
During this work, N.~M. and S.~P. were supported by the CNRS grant number
11240 within the 2002 CNRS/SNF cooperation program. 
P.~H. was employed by Nordita, Copenhagen until 31 July 2002.
This work was completed during a visit to Nordita, and has been partially
supported also by the European Commission under contract HPRN-CT-2000-00130.

\end{document}